\documentclass[aps,twocolumn,prd,showpacs]{revtex4}
\usepackage{graphicx}
\usepackage{amsmath}
\usepackage{amssymb}

\begin{document}

\title{Compatibility of the chameleon-field model with fifth-force experiments, cosmology, 
and PVLAS and CAST results}

\author{Philippe Brax} \email{brax@spht.saclay.cea.fr}
\affiliation{Service de Physique Th\'eorique, Commissariat \`a
  l'\'Energie Atomique--Saclay, 91191 Gif-sur-Yvette Cedex, France}

\author{Carsten van de Bruck} \email{c.vandebruck@sheffield.ac.uk}
\affiliation{Department of Applied Mathematics, University of Sheffield,
Hounsfield Road, Sheffield S3 7RH, United Kingdom}

\author{Anne-Christine Davis} \email{a.c.davis@damtp.cam.ac.uk}
\affiliation{Department of Applied Mathematics and Theoretical
  Physics, Center for Mathematical Sciences, University of Cambridge,
  Wilberforce Road, Cambridge CB3 OWA, United Kingdom}

\date{\today}

\begin{abstract}
We analyse the PVLAS results using a chameleon field whose
properties depend on the environment. We find that, assuming a
runaway bare potential $V(\phi)$ and a universal coupling to
matter, the chameleon potential is such that the scalar field can
act as  dark energy. Moreover the chameleon field model is
compatible with the CAST results, fifth force experiments and
cosmology.
\end{abstract}
 \maketitle

One plausible explanation for the observed accelerated expansion
of the universe is the presence of a pervading scalar field whose
dynamics lead to an approximately constant energy density today
\cite{quint}. As a result, the mass of this scalar field turns out
to be extremely small, i.e. of the order of the present Hubble
rate. Such an almost massless scalar field is in direct conflict
with gravitational experiments when its coupling to matter is of
order of gravitational strength. Indeed  fifth force experiments
give stringent bounds on the gravitational coupling. One must
therefore either decouple almost massless scalar fields from
ordinary matter or shield macroscopic bodies. The former mechanism
has been used to argue that the dilaton of string theory in the
strong coupling regime does not lead to gravitational problems and
can drive the acceleration of the expansion \cite{Damour}. The
latter possibility is at play when scalar fields behave as
chameleon fields. Such fields couple to matter strongly and
non-linear effects can reduce the interaction range of the force
mediated by the chameleon field created by a massive body
\cite{cham1}. This is all due to the presence of a thin-shell
effect whereby the scalar field is essentially constant inside the
massive body, except for a thin shell whose width governs the
strength of the scalar interaction with other massive bodies. For
lighter bodies, the thin shell effect does not appear and
chameleon fields can become invisible provided their environment
dependent mass is large enough.

The coupling of a chameleon field to the electromagnetic sector could lead to
variations of the fine structure constant. In a cosmological
setting, the chameleon field settles down at the bottom of its
time-dependent (effective) potential very early in the universe,
thus preventing large mass variations during Big Bang Nucleosynthesis.
As the minimum of the potential evolves adiabatically with the time variation of the matter
energy density, a variation of the fine structure constant could be
induced, albeit negligible for an order one gravitational coupling \cite{us}.

The coupling of a scalar field to photons has been invoked in
order to explain the original PVLAS dichroism
detection \cite{PVLAS}, recently superseded by measurements
showing upper bounds on both the dichroism and the
birefrigence \cite{newPVLAS}. The scalar field coupling strength
must be suppressed by a scale bounded by  $M\ge 10^6$ GeV for
masses which are typically $m\le 10^{-3}$ eV. It is tantalising
that such a scalar field mass is within the ball park of the
energy density of the universe. Moreover the coupling to photons
\begin{equation}
-\frac{1}{4} \int d^4 x e^{\phi/M} F_{\mu\nu}F^{\mu\nu}
\end{equation}
with $\phi\ll M$ is reminiscent of the coupling of the dilaton to
photons. The former results obtained by the PVLAS collaboration are in
conflict with astrophysical bounds such as CAST \cite{CAST}, which
for the same mass for the scalar field, require much smaller
couplings ($M>10^{10}$GeV).  Recently, a lot of work has been done
in order to explain the discrepancy theoretically (see e.g.
\cite{PVLAStheory}). If a future PVLAS type measurement indicated
a coupling in the range $10^6 {\rm GeV} \le M\le 10^{10} {\rm GeV}$, the 
question of whether this could be made compatible with the CAST result would
be raised. We will address this question in the following.

In this letter we point out that the  PVLAS results in the interesting range  
$10^6 {\rm GeV} \le M\le 10^{10} {\rm GeV}$ with a mass $m\approx 10^{-3}$~eV 
are not in conflict with astrophysical bounds such as CAST if the particle 
concerned were a chameleon. Moreover, low values of $M\approx 10^{6} {\rm GeV}$
are favoured in order to be compatible with dark energy. We require that the 
scalar field couples to all matter forms and, in the following, we will assume
that all the couplings of $\phi$ to matter are universal and are
given by the one suggested by the analysis of the PVLAS experiment.
Our model  is of the scalar-tensor type
\begin{eqnarray}
S&=&\int d^4x \sqrt{-g}\left(\frac{1}{2\kappa_4^2}R-
g^{\mu\nu}\partial_\mu\phi \partial_\nu \phi -V(\phi)
-\frac{e^{\phi/M}}{4} F^2\right)\nonumber \\ &+& S_m( e^{\phi/M}
g_{\mu\nu},\psi_m)
\end{eqnarray}
where $S_m$ is the matter action and the fields $\psi_m$ are the matter
fields. As a consequence, particle masses in the Einstein frame
become
\begin{equation}
m(\phi)= e^{\phi/M} m_0
\end{equation}
where $m_0$ is the bare mass as appearing in $S_m$. The effective
gravitational coupling is given by
\begin{equation}
\beta= \frac{m_{\rm Pl}}{M},
\end{equation}
and therefore very large ($\beta \le 10^{13}$) when assuming the
results from the PVLAS experiment ($M\ge 10^6$~GeV) \cite{newPVLAS}. To prevent
large deviations from Newton's law one must envisage non--linear
effects shielding massive bodies from the scalar field. One
natural possibility is that the scalar field $\phi$ coupled to
photons has a runaway (quintessence)--potential leading to the
chameleon effect. For exponential couplings, this is realised when
\begin{equation}\label{poti}
V(\phi)= \Lambda^4\exp (\Lambda^n/\phi^n) \approx \Lambda^4 +
\frac{\Lambda^{4+n}}{\phi^n}
\end{equation}
The first term corresponds to an effective cosmological constant while the
second term is a Ratra-Peebles inverse power law potential.
Acceleration of the universe is obtained provided $\Lambda\approx
10^{-12}$ GeV, which of course assumes that the scalar field
$\phi$ is responsible for the acceleration of the universe.

In the presence of matter, the dynamics of the scalar field is
determined by an effective potential
\begin{equation}
V_{\rm eff }(\phi)=\Lambda^4\exp (\Lambda^n/\phi^n)+ e^{\phi/M}\rho
- \frac{e^{\phi/M}}{2}\left({\bf E}^2 - {\bf B}^2 \right)
\end{equation}
where $\rho$ is the energy density of non-relativistic matter and we have used the fact
that $(1/4)F_{\mu\nu}F^{\mu\nu} = -({\bf E}^2 - {\bf B}^2)/2$, with ${\bf E}$ and ${\bf B}$ being
the electric and magnetic field, respectively. The origin of the last terms is due
to the coupling between matter, electromagnetism and the scalar field. Since we will consider
the PVLAS experiment in the following, we will set ${\bf E} = 0$.

The effective potential
leads to a stabilisation of the scalar field for
\begin{equation}\label{minimum}
\phi= \left(\frac{n \Lambda^{4+n}M}{ \rho_{\rm tot}}\right)^{1/(n+1)},
\end{equation}
where $\rho_{\rm tot} = \rho + {\bf B}^2/2$ is the effective energy density with contributions from
both matter and the magnetic field. The mass at the bottom of the potential is given by
\begin{equation}
m^2= n(n+1) \frac{\Lambda^{n+4}}{\phi^{n+2}}
\end{equation}
Let us now consider the case of the vacuum chamber used in the
PVLAS experiment such that the energy density inside the cavity is
$\rho_{\rm tot}$ and the mass of the scalar field $m_{\rm lab}$,
then
\begin{equation}\label{relat}
\Lambda^{4+n}= \frac{(n+1)^{n+1}}{n}  \rho_{\rm tot}^{n+2} M^{-n-2}
m_{\rm lab}^{-2n-2}
\end{equation}
 Taking $m_{\rm lab}= 10^{-3}$~eV, considering the fact that the
density has contributions from the gas ($\rho_{\rm lab,gas}\approx
2\times 10^{-14}$~g/cm$^3$) and the magnetic field ($B=5$~T,
corresponding to $\rho_{\rm lab,field} \approx 7 \times
10^{-14}$~g/cm$^3$), and the lower bound $M = 10^6$ GeV,
determines $\Lambda$
\begin{equation}
\Lambda^{4+n} \approx \frac{(n+1)^{n+1}}{n}  10^{-12n - 48}
\end{equation}
For  $n={\cal{O}}(1)$ we find that
\begin{equation}
\Lambda \approx 10^{-12} \rm GeV
\end{equation}
as required to generate the acceleration of the universe. Hence we
find that the cosmological constant $\Lambda^4$ is compatible with
the laboratory experiments. Higher values of the coupling scale
$M$ and the mass $m_{\rm lab}$ would lead to a smaller value of
$\Lambda$, incompatible with cosmology. Hence we only consider the
lower bound $M=10^6$ GeV in the following.

As already mentioned, the lower bound given by the PVLAS
experiment is in conflict with the CAST experiment on the
detection of scalar particles emanating from the sun, as it
requires $M\ge 10^{10}$ GeV. However, this bound does not apply
when the mass of the scalar field in the sun exceeds $10^{-5} \rm
GeV$. Let us evaluate the mass of the chameleon field inside the
sun. Using eqn. (\ref{relat}) one obtains
\begin{equation}\label{relati}
m_{\rm sun}= m_{\rm lab} \left(\frac{\rho_{\rm sun}}{\rho_{\rm
lab}}\right)^{(n+2)/2(n+1)}.
\end{equation}
Now $\rho_{\rm sun}/\rho_{\rm lab}\approx 10^{14}$ and, with
$n=0(1)$, one finds
\begin{equation}
m_{\rm sun} \sim 10^{-2} {\rm GeV} \gg 10^{-5} {\rm GeV}
\end{equation}
implying no production of chameleons deep inside the sun.
Chameleon particles can also be produced at the surface of the sun where
the density $\rho_{out}$ is much lower. Taking
$\rho_{\rm out}/\rho_{\rm lab}\approx 10^{9}$, we find that the CAST results
can be explained when $n\le 1$. Hence, the CAST
experiment is in agreement with the chameleon model due to
the fact that the chameleon field is very massive in the sun.

The PVLAS experiment puts constraints on the mass of the scalar
field inside the field zone and the coupling constant $M$. Since
the mass of the chameleon field depends on the ambient energy
density, the theory predicts that the results of the PVLAS
experiment would have been different if the density of matter
inside the field zone and the magnetic field strength were
different.  Since the amplitude of the dichroism depends on
$m_\phi^{-4}$ \cite{theory,PVLAS}, we would expect that, from eq.
(\ref{relat}), the amplitude of dichroism would decrease as
$\rho_{\rm gas}$ increases (if all other parameters are fixed). In
our theory, the amplitude of dichroism depends in a very
non-trivial way on both the pressure and the strength $B$  of the
magnetic field. Observing such variations would be a test of the
theory described here. If the magnetic field strength is kept
fixed at $B=5.5$~T and decreasing $\rho_{\rm lab,gas}$, according
to our theory there would be a saturation soon, since the total
energy density inside the field zone is dominated by $\rho_{\rm
lab,field}$ and the chameleon mass becomes independent of
$\rho_{\rm lab, gas}$. If the density of gas is further increased,
the mass of the chameleon field changes and therefore the
amplitude of the observed effect. Fixing the gas density but
changing $B$ would result in a non-trivial dependence of the
amplitude of the dichroism on $B$, since the mass of the chameleon
depends on $B$. The detailed analysis of these effects is in
progress \cite{us1}.

Let us now analyse the gravitational tests on earth and in the
solar system. As we will now argue, current experiments will
not be able to detect a new force mediated by the chameleon
field, even if the coupling $\beta$ is large. The argument
is as follows:

On earth, and for experiments performed in the atmosphere where the
density is $10^{-3}$ $g/cm^3$, the mass of the chameleon field is
$m_{\rm atm}\approx 10^{-5}$ GeV leading to a very short-ranged
interaction, hence not detectable in gravity experiments.
Similarly in the solar system where $\rho_{\rm solar}= 10^{-24}
g/cm^3$, the mass of the chameleon becomes $m_{\rm solar}=
10^{-22} $ GeV, with a range of $10^6 $ m, too small to affect the
motion of planets. Finally let us consider the satellite gravity
experiments. As the range of the chameleon force in the galactic
vacuum is much larger than the size of a satellite, one would
expect large deviations from Newton's law for satellite
experiments. This is only the case if the thin shell mechanism is
not at play. A test mass of a gedanken experiment aboard a
satellite has a thin shell provided $\phi_{\rm solar} \le \beta
\Phi_N$ where $\Phi_N$ is Newton's potential at the surface of the
test body. For a typical test body of mass 40 g and radius 1 cm,
Newton's potential is $\Phi_N \approx 10^{-27}m_{Pl}$ implying
that a thin shell exists for $\phi_{\rm solar} \le 100 $ GeV. We
find that
\begin{equation}
\phi_{\rm solar} \approx 10^{-(1+12n)/(n+1)} \ll 10^2
\end{equation}
implying that satellite experiments would not detect any deviation
from Newton's law due to a force mediated by the chameleon field.

A detailed analysis of theories with scalar fields strongly coupled
to matter has been carried out in \cite{Mota}. A range of different
constraints have been investigated and, for the model at hand,
current constraints coming from local experiments are fulfilled. The large
coupling in the theory implied by the PVLAS experiment ensures
that  the chameleon effect is very efficient, resulting in a large
effective mass for the scalar field and, consequently, short interaction
range of the force mediated by the chameleon.

For $\beta\gg 1$, Casimir force experiments provide tight bounds
on the model parameter. For large $\beta$, it was found that the
scale $\Lambda$ cannot be much larger than that set by the
cosmological constant $\Lambda = 10^{-3}$eV \cite{Mota}. However,
current experiments are compatible with the parameter $M$ and
$\Lambda$ in our model.
More precisely, the ratio of the chameleon force over the Casimir force for a two plate geometry
is given by \cite{Mota}
\begin{equation}
\frac{F_{\phi}}{F_{\rm Cas}}= \frac{240}{\pi^2} K_n (\Lambda d)^{\frac{2(n+4)}{n+2}}
\end{equation}
where $K_n$  is expressed in terms of Euler's beta function
$K_n=(\sqrt 2 \frac{B(1/2,1/n+1/2)}{n})^{2n/(n+2)}$ and
$\frac{240}{\pi^2} K_n\approx 40$  for $n\le O(1)$; $d$ is the
interplate distance. Casimir forces have been measured up to $d=
10$ $\mu$m implying that $\Lambda d \le 0.1$ and therefore
$F_{\phi}/F_{Cas}\le 10^{-2}$. Such an accuracy is below the
present experimental levels. Detailed work on Casimir constraints
is in progress \cite{us2}.

Astrophysical constraints, such as those coming from neutron stars and white
dwarfs also constrain the existence of strongly coupled scalar fields. The
force mediated by $\phi$ could alter the stability of such stars. However,
the constraints coming from these considerations are not very strong and for
the parameter at hand the theory is compatible with observations.

On scales relevant for cosmology, the chameleon mediates a force which
could affect structure formation.  The interaction range is given by
$\lambda = V_{,\phi\phi}^{-1/2}$, below which the effective gravitational
constant is $G_{\rm eff} = G_{\rm N}\left[1+2\beta^2\right]$. For
the potential (\ref{poti}), the interaction range is (assuming $\Lambda = 10^{-3}$eV)
\begin{equation}
\lambda_{\rm cham} \approx 10^{-2}\left(\frac{\phi}{\Lambda}\right)^{1+n/2} {\rm cm}.
\end{equation}
As an example, with $n=1$, eq. (\ref{minimum}) gives $\phi =
10^{-4}$GeV in the minimum today, so that, for $\Lambda =
10^{-3}$eV, $\lambda_{\rm cham} = 10^{10}$~cm, which is of the
order of the radius of the sun and hence cosmologically
irrelevant. We remark that the interaction range is much larger if
$\beta$ is smaller: $\lambda_{\rm{cham}} \approx 100$~pc for
$\beta={\cal O}(1)$\cite{us}. As is the case with local
experiments, cosmologically the chameleon mechanism is more
effective for large $\beta$.

As analysed in \cite{us}, the chameleon must be stuck at
the bottom of the potential since before BBN. Thus the chameleon has
the following evolution
\begin{equation}
\phi_{\rm cos} \approx (\Lambda^{n} M)^{1/(n+1)} (1+z)^{-3/(n+1)}
\end{equation}
which is a valid approximation as long as $\phi \ge \Lambda$. In
particular, the fine-structure constant is such that
\begin{equation}
\frac{\alpha^{-1}(z)- \alpha^{-1}(0)}{\alpha^{-1}(0)}=
\left(\frac{\Lambda}{M}\right)^{n/(1+n)}\left((1+z)^{-3/(n+1)}-1\right) \end{equation}
The prefactor is very small ($10^{-8}$ for $n=1$ and smaller for larger $n$).
This implies that the variation of the fine structure constant is negligible
compared to the results presented in \cite{murphy}.

Hence the chameleon field is not observable either in current
gravitational experiments or cosmologically. In addition to its role
in the PVLAS experiment, the chameleon Lagrangian possesses terms
like
\begin{equation}
\lambda \frac{H\phi}{M} \psi \bar \psi
\end{equation}
coupling two fermions, one Higgs field and the chameleon field.
After electroweak symmetry breaking when the Higgs field picks up
a vev, the effective coupling becomes
\begin{equation}
\frac{m_\psi}{M} \phi \bar \psi \psi \label{coupling}
\end{equation}
so the chameleon couples like an almost massless Higgs boson to
the standard model fermions. The only difference is that the Higgs
coupling is suppressed by the electroweak vev $v$. The weakness of
the chameleon coupling is measured by the ratio $v/M\sim 10^{-4}$.
Such a small coupling makes the chameleon detection unrealistic
even at LHC scales.

However, such a Yukawa coupling could lead to deviations from standard
model results in high precision experiments. For example, the
anomalous magnetic moment of the muon (or the electron) could be
affected. New contributions occur due to the chameleon coupling to
fermions, as in eqn. (\ref{coupling}). These can be evaluated
by replacing photon lines with chameleon lines in the relevant
Feynman diagrams. This gives a contribution to $(g-2)$ of
$(m_{e,\mu}/M)^2$, which is of order $10^{-12}$ for the muon,
instead of the usual $\alpha_{QED}$. It is thus suppressed.
Another contribution comes from hadronic loops which is about
two orders of magnitude larger than that of the chameleon (see
e.g.~\cite{Davier} and references therein for a recent discussion).
Hence the effect on the anomalous magnetic moment is negligible.
Similarly, the couplings in eqn. (\ref{coupling}) could lead
to corrections to the hyperfine structure of the hydrogen atom.
As for $(g-2)$, the effect spring from one loop contributions
obtained by replacing photon propagators by those for chameleons.
Hence, the effect on the energy levels is of order $(m_e/M)^2$ compared
to $\alpha_{\rm QED}$, and is thus negligible.

In conclusion, a scalar field strongly coupled to matter, with
coupling strength as suggested by  the new  PVLAS results, and
non-linear self--interactions, is compatible with current
fifth-force experiments and cosmology. Moreover a coupling strength with $M\approx 10^{6} {\rm GeV}$ is favoured
when considering compatibility with dark energy. In such a range, 
chameleons provide a natural
explanation for the discrepancy with the CAST results, since the
(effective) mass of the scalar field depends on the matter density
of the environment. Future laboratory experiments, such as Casimir
force experiments, could be designed to detect the force mediated
by the scalar field and would be an independent test from the
PVLAS experiment. We have pointed out that, according to the model
discussed in this paper, the amplitude of the effect in the PVLAS
experiment depends on the mass density and the magnetic field
inside the field zone. Therefore it would be interesting to
investigate both experimentally and theoretically whether there is
any dependence on both the density of ambient matter and the
magnetic field. Work on this topic is in progress \cite{us1}.

\acknowledgements We thank G.~Cantatore, C.~Rizzo and D.~Shaw for helpful discussions.
ACD thanks CEA Saclay for their hospitality. CvdB and ACD thank PPARC for
partial support. PhB acknowledges support from RTN European programme
MRN-CT-2004-503369.

\end{document}